\documentclass[11pt,twoside]{article}

\usepackage{asp2006}
\usepackage{epsf}
\usepackage{psfig}
\usepackage{lscape}
\usepackage{graphicx}

\begin{document}
\title{Non-thermal radio emission from OB stars: an observer's view}   
\author{Paula Benaglia$^{1,2}$}   
\affil{\affil{%
(1) Instituto Argentino de Radioastronom\'\i a, CCT La Plata-CONICET, C.C. 5, 1894 Villa Elisa, Argentina\\
(2) Facultad de Ciencias Astron\'omicas y Geof\'\i sicas, Paseo del Bosque s/n, 1900 La Plata, Argentina; pbenaglia@fcaglp.unlp.edu.ar\\}}    

\begin{abstract} Some early-type stars are detectable radio emitters; their spectra can present both thermal and non-thermal contributions. Here I review the public radio data on OB stars, focusing on the non-thermal sources. The analysis of the statistical results gives rise to many open questions that are expected to be addressed, at least in part, with the upgrades of current radio telescopes and the upcoming new generation instruments.
\end{abstract}

\section{Introduction}  

Early-type stars like OB stars and their descendants Wolf-Rayet (WR) stars are characterized by strong winds that continuously expel matter from the stellar atmospheres. The velocity of the outflow can reach thousands of km s$^{-1}$, and the mass is lost at rates up to 10$^{-4}$ M$_\odot$ yr$^{-1}$. The plasma forming the winds is optically thick and radiates at radio continuum by thermal Bremsstrahlung. The winds are prone to also suffer instabilities that give rise to shocks. In massive binary systems, regions where winds from the two stars collide are permeated by even stronger and larger shocks, which, in turn, accelerate particles up to relativistic energies, and non-thermal (synchrotron) emission is generated.\\

In the case of uniform, homogeneous winds, the measurement of the thermal radio-flux density at a given frequency allows the estimate of an average value of the stellar mass loss rate, provided some basic wind parameters are known, and the stellar distance is determined. The result will be an overestimation if there are inhomogeneities in the plasma or if the measured flux density has contribution of thermal and  non-thermal emission. The velocity at which a massive star loses mass is a fundamental input in stellar evolution and population models, and slightly different values can change the results drastically.\\

Radio observations at more than one frequency - preferentially, along the spectrum - are important to describe the emission regime, to disentangle the different contributions to the flux, and to study the emission mechanisms that lead to the observed spectrum. 
The detection of synchrotron emission is by itself an evidence for the existence of relativistic particles. These particles, by interactions with other ones and with magnetic and photon fields are also involved in  the production of high-energy radiation\footnote{A key point is whether the associated high-energy radiation will be detected by the present telescopes.}.
Thus, the study of the radio emission from the stellar winds can give information on the population of relativistic leptons in the winds, 
and about the possible non-thermal high energy emission from massive stars (De Becker 2007). 
Moreover, polarization measurements at non-thermal stellar sources can provide insights on the magnetic field of massive stars, one of the stellar parameters most difficult to measure.\\

In the last decades, hundreds of OB stars have been observed, many of them at more than one wavelengths. A study of the detected cases is presented here, based mainly on a statistical analysis. Examples of instrumental developments, in either existent or new facilities, that will easily allow to enlarge the number of detections by more than an order of magnitude closes this review.  

\section{Milestones in stellar wind studies}    

It could be said that the history of the studies on stellar winds begun with the discovery of SNe 3c10, the ``Tycho'' supernova, in 1572, and continued with the monitoring of subsequent novae, supernovae, and LBVs magnitude changes. With time, as spectroscopy became a powerful tool, C. Wolf and G. Rayet (1867) discovered that a few stars showed unusually broad emission line spectra. P-Cygni profiles were measured (Campbell 1892), and they were interpreted by taking into account Doppler shifts, revealing signatures of outflows.\\

It was early in the XXth Century that the hypothesis that selective absorption can drive the outflows was formulated (Milne 1924, Johnson 1925). L. Biermann, among others, proposed the existence of a continuum solar ejecta and expanding envelopes, and also discovered the first evidence of that, through the deflection of a comet tail (Biermann 1951). Surprisingly, it was not until 1958-60, that the expressions `solar wind' and `stellar wind' were coined (Parker 1958, 1960). By those years, the OB-winds studies greatly benefited from the progress and knowledge that were being gained on the solar wind. 
In the year 1967 the Mariner 2 probe made the first detection of the solar wind. In the same year the first OB-UV spectra were measured by D. C. Morton using a detector on a rocket. Breakthroughs began to accumulate in the seventies: Lucy \&  Solomon (1970) developed a  'radiation-driven wind model', radio observations toward massive stars were started, the CAK (Castor, Abbott \& Klein 1975) theory burst in and good predictions on mass loss rates and terminal velocities became available. Wright \& Barlow (1975) and Panagia \& Felli (1975) proposed a thermal description for the emission of stellar winds. The first image of a resolved wind was published  
in 1989 (Moran et al. 1989), and soon VLBI mapping, phase monitoring, and multifrequency studies on stellar winds begun.\\

\section{Stellar wind regions for particle acceleration}

In a system composed by two stars with winds there are some regions with shocks that are capable to accelerate particles up to very high energies, namely, the wind of each star is prone to suffer instabilities that produce shocks, the region where single supersonic winds encounters the interstellar matter, the region where the two winds collide (see Figure 1). The latter region has a rather strong magnetic field, related to the stars. In such a scenario non-thermal emission can be produced. The same population of relativistic particles will be involved in processes that also produce high-energy photons. Electrons accelerated in the electrostatic field of the ions give rise to relativistic Bremsstrahlung emission. The copious UV photons from the OB stars are boosted by relativistic electrons to higher frequencies (IC scattering). Synchrotron emission is produced by electrons accelerated at the existent magnetic field. Neutral pions generated by inelastic collisions between relativistic and thermal protons decay into gamma rays. The possibility that high energy emission can also be produced at the environs of early-type stars motivates the search for counterparts to the unidentified gamma-ray sources, e.g. Benaglia et al. (2005), Romero et al. (1999).

\begin{figure}[!ht]
\begin{center}
\includegraphics[bb=0 0 256 218,width=7cm]{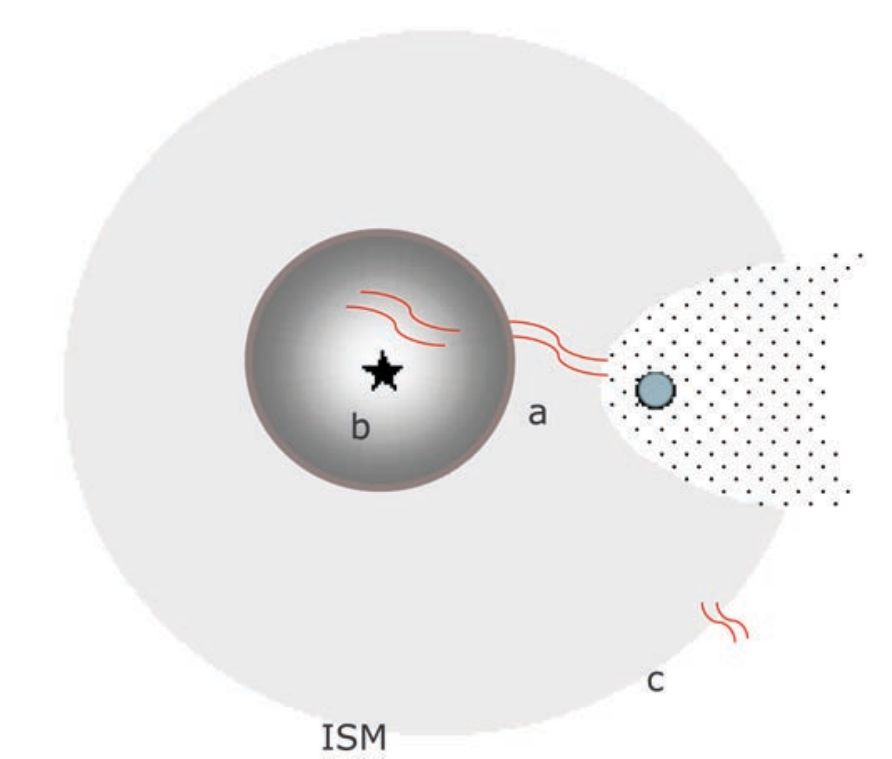}
\end{center}
\caption{Possible regions for shock-mediated particle acceleration in massive star winds: region a) where, in a massive binary system, two stellar winds collide; region b) in the unstable single stellar wind of an early-type star; region c) at the terminal shock produced when and where the stellar wind encounters the ISM.}\label{fig1}
\end{figure}

\section{Published radio continuum data on OB stars}


In 1995, a catalogue of all observed stars at radio wavelengths till that date was presented by Wendker (1995). It listed information on 3021 stellar objects with spectral types from O to M. About 275 were cataloged as O $-$ B2 stars, of which around 40 were detected, at one or more radio frequencies. The works cited by Wendker on the detected cases were reviewed in order to derive spectral index values. The bibliographic sources consulted were Gibson \& Hjellming (1974), Abbott et al. (1980, 1984), Drake et al. (1987), Persi et al. (1988), Bieging et al. (1989), Drake (1990), Leitherer \& Robert (1991), and Howarth \& Brown (1991), Dougherty (1993).\\

A search of OB stars radio observations in the literature since 1995 to date yields the papers by Leitherer et al. (1995), Contreras et al. (1997), Scuderi et al. (1998), Benaglia et al. (2001), Rauw et al. (2002), Blomme et al. (2002, 2003), Setia Gunawan et al. (2003a, 2003b), Benaglia \& Koribalski (2004), De Becker et al. (2004, 2005), Blomme et al. (2005), Puls et al. (2006), Benaglia et al. (2006), Blomme et al. (2007), Benaglia \& Koribalski (2007), Benaglia et al. (2007), Schnerr et al. (2007), Petr-Gotzens \& Massi (2007), van Loo et al. (2008), and Dougherty \& Kennedy (2009).\\
 
Table 1 lists the stars detected at least at one frequency, their spectral type, and references. A spectral index was derive whenever possible. The listed stars are divided in three groups according to the spectral indices, labeled as `Uncl', `NT', and `T'. When a spectral index $\alpha$ is $> +0.3$ ($S_\nu \propto \nu^\alpha$) the source is classified as `Th' (thermal); if $\alpha \rightarrow 0.0$ or less, the source is `NT' (non-thermal). For stars detected at only one frequency, the spectral index remains unknown, or unclassified (`Uncl').  
The third column quotes the reference from where the radio data were gathered. It is usually the most recent reference, which refers to previous ones. It does not necessarily mean that the observations were carried out by the authors of the paper. An entry (NT) means ``probably NT''. It is classified in this way because either it is suggested at the reference quoted, or it was detected at one frequency, and the detection limit at a longer frequency is lower than the flux detected at the shorter frequency. A note (Th) means ``probably thermal'', as proposed by the corresponding reference, or it was detected at one frequency, and the detection limit at a shorter frequency is lower than the flux detected at the longer one.\\

The spectral types were mostly taken from the GOS Catalogue (Ma\'{\i}z-Apell\'aniz et al. 2004), except when otherwise notice. Spectral information on binarity, from Mason et al. (2009), is given in brackets.

\begin{table}[!ht]
\caption{O -- B2 stars with detected radio emission from winds.}
\smallskip
\begin{center}
{\tiny
\begin{tabular}{l l l l c c l}
\tableline
\noalign{\smallskip}
Name       & Sp.Class    & Ref. & Observed $\nu$  & Sp. Index & $d$ & Status \\
           &             &      &  [GHz]          & group     &[kpc]&        \\
\noalign{\smallskip}
\tableline
\noalign{\smallskip}
HD 93129A	 &	O2If*+O3.5V	    &	a &	1.4 -- 24.5	    &	 NT	 &	2.5	&	Bin\\
HD 150136	 &	O3.5If*+O6V+..	&	a & 1.4,2.4,4.8,8.6	&	 NT	 &	1.4	&	Mult,[SB10]\\
HD 93250 	 &	O3.5V((f+))	    &	b	&	4.8,8.6			    &	(NT) &	2.2	&	[C]\\
HD 66811   &	O4 I(n)f	      &	c & 1.4,4.8,8.5,15	&	 Th	 &	0.4	&	[C]\\ 
HD 190429A &	O4 If+	        &	d & 8.5	            &	Uncl &	1.7	&	\\
HD 16691	 &	O4I f+ 	        &	e	& 4.9							&	(Th) &		  &	\\
HD 15570   &	O4If+	          &	f &	4.8,8.4,1.3mm		&	(Th) &	2.2	&	\\
HD 164794	 &	O4((f))+?	      &	g & 1.4,4.8,15		  &	 NT	 &	1.6	&	[SB2?]\\ 
CD-47 4551 &	O5If	          &	a	& 1.4,2.4,4.8,8.6	&	 NT	 &	1.7	&	[C]\\
HD 14947	 &	O5If+	          &	d &	4.9,8.5			    &	(Th) &	2	  &	Cte\\
Cyg OB2 \#7&	O5If+	          &	h	&	1.4,2.4,4.8,8.6 &	Uncl &	1.7	&	\\
Cyg OB2 \#9&	O5If+ 	        &	i	&	1.4,4.8,8.4,15	&	 NT	 &	1.7	&	Bin,[SB2?]\\
HD 15558	 &	O5III(f)+O7V	  &	j	&	4.8					    &	(NT) &	2,2	&	Bin,[SB20]\\
Cyg OB2 \#11&	O5III(f)  	    &	j	&	1.4,4.8					&	(NT) &	1.7	&	[SB1?]\\
HD 168112	 &	O5.5III(f*)	    &	k	&	4.8,8.5,15			&	 NT	 &	2	  &	Bin,[SB1?]\\ 
HD 108		 &	O6:f?pe	        &	e	&	4.9							&	(Th) &	2.5	&	[C]\\
HD 210839  &	O6I(n)fp        &	l &	1.4,8.4		      &	 Th	 &	0.8	&	Cte\\  
Cyg OB2 \#8A&O6Ib(n)(f)+O5.5III(f)&h&.4,2.4,4.8,8.6	&	 NT	 &	1.7	&	Bin\\
HD 124314	 &	O6V(n)((f))	    & a	&	1.4,2.4,4.8,8.6	&	 NT	 &	1	  &	[SB1?]\\
HD 206267	 &O6.5V((f))+O9.5:V	&	m	&	4.9	        		&	Uncl &	0.8	&	Bin\\
HD 167971	 &	O6.5V+O5-8V	    & n	&	1.4,4.9,8.5			&	 NT	 &	2	  &	Mult,[SBE]\\
HD 150135	 &	O6.5V((f))	    &	o	& 4.8,8.6				  &	 Th	 &	1.4	&	$^{\dag}$\\
HD 152623	 &	O7V(n)((f))	    &	p	&	1.4,2.4,4.8,8.6	&	 NT	 &	1.9	&	[SB1O]\\
HD 47839   &	O7V((f))+O9.5V	&l,m& 1.4,4.9,8.5			&	 NT	 &	0.8	&	Bin,[SB10]\\ 
Cyg OB2 335&	O7V	            & h &	1.4,2.4,4.8,8.6 &	 NT	 &	1.7	&	\\
HD 166734  &	O7Ib(f)	        & j	&	4.9 						&	(Th) &		  &	\\
Cyg OB2\#5 &	O7Ia+Ofpe/WN9	  &	q	&	4.9,8.5,22			&	 NT	 &	1.7	&	Mult,[SB2OE]\\
HD 24912   &	O7.5III(n)((f))	&	l	& 1.4,8.4				  &	 Th	 &	0.5	&	Cte\\ 
HD 47129	 &	O7.5I+O6I	      &	e	&	4.9							&	(Th) &	1.5	&	\\
HD 151804	 &	O8Iaf	          &	j	&	4.9,15			    &	 NT	 &	1.9	&	Cte\\
HD 19820   & O8.5III((n))+B0V	&	r	& 2.7,8				    &	(NT) &	1	&	Bin\\ 
HD 37043	 &	O9III+B7IV	    & s	&	8.4			        &	Uncl &	0.5	&	\\
HD 57061   &	O9II+B2V	      &	s	& 8.4				      &	Uncl &	1.5	&	\\
HD 149404	 &	O9Ia+O6.5III	  &	s	&	8.4				      &	Uncl &	1.4	&	\\
HD 76341	 &	O9 Ib	          & v	&	8.4				      &	Uncl &	1.8	&	Cte\\
HD 37742	 &	O9.7Ib+B2III	  &s,t&	4.9,8.4		      &	Th	 &	0.5	&	\\ 
HD 149757	 &	O9.5Vnn	        & s &	8.4           	&	Uncl &	0.2	&	Cte\\ 
HD 37468	 &	O9.5V+B0V+B2..	&	m	& 4.9,8.5,15		  &	NT	 &	0.5	&	Mult,[SB2?]\\
HD 36486 	 &	O9.5II+B0III   	&	s	&	8.4				      &	(NT) &	0.5	&	Bin,[SB10E]\\ 
HD 209975	 &	O9.5Ib	        &	u	&	15			        &	Uncl &	0.8	&	\\
Cyg OB2 \#10&	O9.5Ib	        &	u	&	15			        &	Uncl &	1.7	&	\\
HD 30614   &	O9.5Ia	        &d,l&	1.4,4.9,8.5,15  &	 Th	 &	1	  &	Cte\\ 
HD 195592	 &	O9.5Ia	        & d	&	4.9,8.5,15			&	Th	 &	1.3	&	SB1?\\
HD 152424	 &	CO9.7Ia	        &	s	& 8.4				      &	Uncl &	1.9	&	SB1?\\
HD 163181	 &	BN0.5Iap	      &	o	& 4.8,8.6				  &	Th	 &	1.4	&	EB$^{\dag\dag}$\\
MWC 349	   &	B[e]+B0III	    &	h	&	1.4,2.4,4.8,8.6	&	Th	 &	1.2	&	Bin\\
HD 37128   &	B0 Ia	          &	w & 1.4,15,20	      &	Th	 &	0.5	&	\\ 
HD 204172	 &	B0 Ib	          &	v	&	8.5				      &	Uncl &	3	  &	\\
HD 154090	 &	B0.7 Ia	        &	v	&	8.5     				&	Uncl &	1.1	&	\\
HD 5394		 &	B0IVe	          &	x	&	8.5				      &	Th	 &	0.25&	\\ 
$\theta$ Ori&	B0.5V+TT+*	    &	z	&	4.9,9.5 				&	NT 	 &		  &	Mult\\
HD 193237	 &	B1 Ia	          &	d &	1.4,5,8.5,15		&	Th	 &	2	&	\\ 
HD 190603	 &	B1.5 Iae        &	d & 4.9,8.5,15	    & NT	 &	2	&	\\
HD 194279	 &	B1.5 Ia	        &	d &	8.5,15			    &	Th	 &	1	&	\\
BD-14 5037 &	B1.5 Ia	        &	d	&	4.9,8.5,15			&	Th	 &	1.7	&	\\
HD 148379	 &	B1.5 Iape	      &	v	&	17 	            &	Uncl &	1.3	&	\\
HD 37017   &	B1.5V	          &	y	&	8.5		          &	(NT) &		&	SB\\
HD 2905		 &	B1Ia	          &	d & 8.5		          &	Uncl &	1.1	&	\\ 
HD 152236  &	B1Ia	          &	p	&	4.8,8.6         &	Th	 &	1.8	&	\\
HD 169454	 &	B1Ia	          &	j	&	4.9,15			    &	Th	 &	0.9	&	\\
HD 36485 	 &	B2 V	          &	y	&	8.5				      &	(NT) &		  &	Bin\\ 
HD 41117	 &	B2Ia	          &	d	&	4.9,8.5,15			&	Th	 &	1.5	&	\\
HD 80077	 &	B2Ia+	          &	b	&	4.8,8.6	        &	Th	 &	3	&	\\
HD 37479	 &	B2Vp	          & y	&	8.5				      &	(NT) &	0.5	&	Mult\\ 
\noalign{\smallskip}
\tableline
\end{tabular}
}
\end{center}
{\small a: Benaglia et al. 2006; b: Leitherer et al. 1995; c: Blomme et al. 2003; d: Scuderi et al. 1998; e: Persi et al. 1988; f: Lamers \& Leitherer 1993; g: Rauw et al. 2002; h: Setia Gunawan et al. 2003a; i: van Loo et al. 2008; j: Bieging et al. 1989; k: Blomme et al. 2006; l: Schnerr et al. 2007; m: Drake 1990; n: Blomme et al. 2007; o: Benaglia et al. 2001; p: Setia Gunawan et al. 2003b; q: Dougherty \& Kennedy  2009; r: Gibson \& Hjellming 1974; s: Howarth \& Brown 1991; t: Abbott et al. 1980; u: Puls et al. 2006; v: Benaglia et al. 2007; w: Blomme et al. 2002; x: Dougherty 1993; y: Drake et al. 1987; z: Petr-Gotzens \& Massi (2007); $\dag$: Barb\'a (private communication); $^{\dag\dag}$Bulut \& Demircan 2007.}
\end{table}

\subsection{Bias and error sources}

To derive the spectral index information for each case, the flux density of the original references were used. 
Because of the extremely high sensitivity needed for measuring continuum emission from stellar winds (below the mJy), and the angular resolution needed to resolve the systems (below the arcsec), almost all detections of OB winds revealed point sources. In this respect,
the information presented here must be taken with caution. The observing campaigns were carried out using different telescopes and even the same telescope was used at different epochs, i.e. different performances and likely different states of the source. Table 1 gathers radio data from 26 studies (see references at Table 1, Column 3). Each group has been selected with different criterion, such as stellar distance and luminosity, celestial regions, spectral type range, declination coverage, etc. Different sensitivities led to fix different noise thresholds, which implies different detection limits. The various angular resolutions used yielded to resolved multiple systems in some cases, and in others a multiple system was considered as a point source; the exact system composition, in many cases, is still unknown.\\

For the studies of binary systems, the epoch of the observation is often critic, since the radio flux is  expected to vary along the orbital phase. This issue could not be taken yet into account when observing OB stars in the radio domain, since radio telescopes with still better sensitivity than that now available and a larger {\sl uv} coverage (especially for southern stars) are needed.\\

Some spectral indices were derived from data taken at different epochs, months or even years apart. This is a potential source of error particularly for massive binary systems where the flux density can be sensitive to the orbital phase. If the system is wide (period: $\sim$ decades) and is away from periastron, the spectral index derived can be taken as a reasonable average. But this is not the case if the system is close or approaching periastron. Even if the observation was set at two separate frequencies simultaneously, it should be considered that the spectral index is valid only at one particular orbital phase. For systems with short periods (hours) a 12-h synthesis observations allow only to derive a spectral index averaged over the whole period.\\

An additional problem introduced when observing near-to-periastron systems is that the non-thermal emission can be absorbed by the thermal one from the single winds, hiding a possible NT case.\\ 

The observations analyzed in the present paper were taken along 35 years: spectral classification of the sources have varied with time, in accordance with the determination of more accurate optical spectra. All stellar parameters derived from the measured flux densities (like the mass loss rate, for instance) are only indicative and can be superseded by further studies.\\


\section{Statistical analysis and results}

As mentioned in the previous Section, the thorough work by Wendker (1995) comprised the results of observations toward 3021 stellar-like sources, 274 of which were catalogued as O-B2 spectral type stars. It can be appreciated than around 440 objects were detected, and about 40 of them were O-B2 stars.\\ 

If the observations taken from 1995 to 2008 are added to those recorded by Wendker (1995), a total number of $\sim$ 65 O-B2 stars have been detected up to date. Around a 25\% of them lacks of information on the radiation regime; for the rest it has been possible to derive spectral indices. Approximately half of the $\alpha$ values are close to $+$0.6, whereas the other half are around or below zero. Figure 2 shows the number of detected stars vs spectral type, and also the information on the spectral index.\\

\begin{figure}[!ht]
\begin{center}
\includegraphics[bb=0 0 467 154,width=11cm]{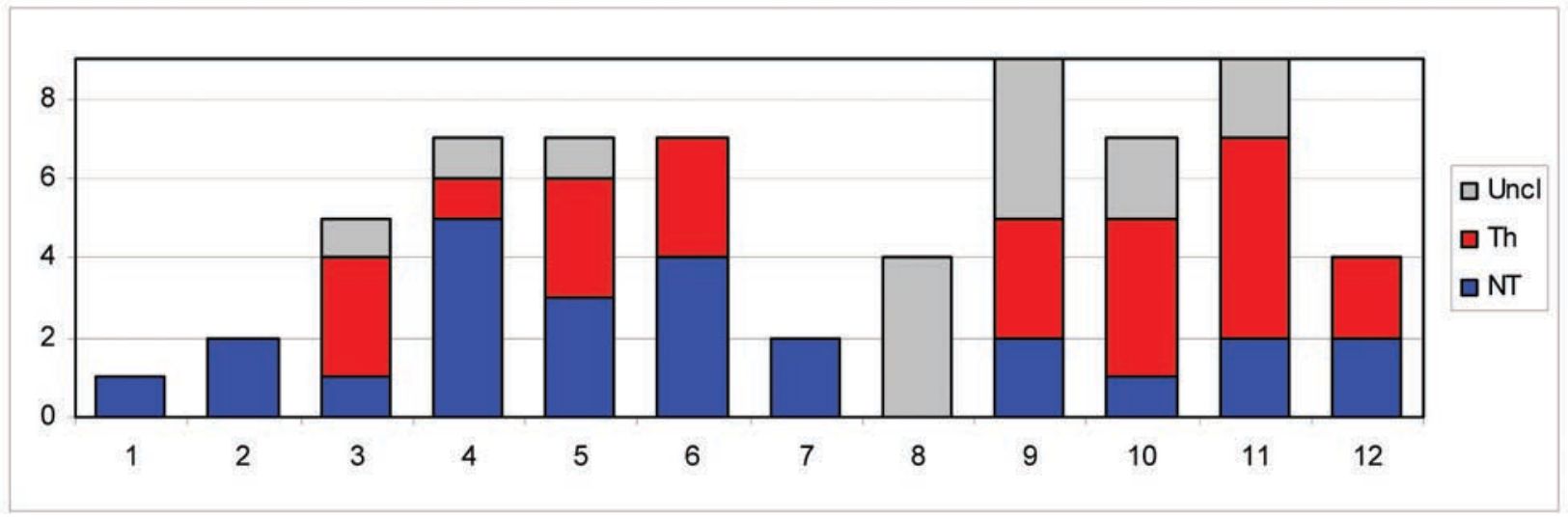}
\end{center}
\caption{Number of detected O--B2 stars as a function of spectral type. Blue bars: stars with spectral indices $\alpha$ equal to 0 or negative (NT). Red bars: stars with $\alpha$ near +0.6 (Th). Gray bars: stars with no spectral index information (Uncl).}\label{fig2}
\end{figure}

\vspace{1.5cm}
\begin{figure}[!ht]
\begin{center}
\includegraphics[bb= 0 0 461 182,width=11cm]{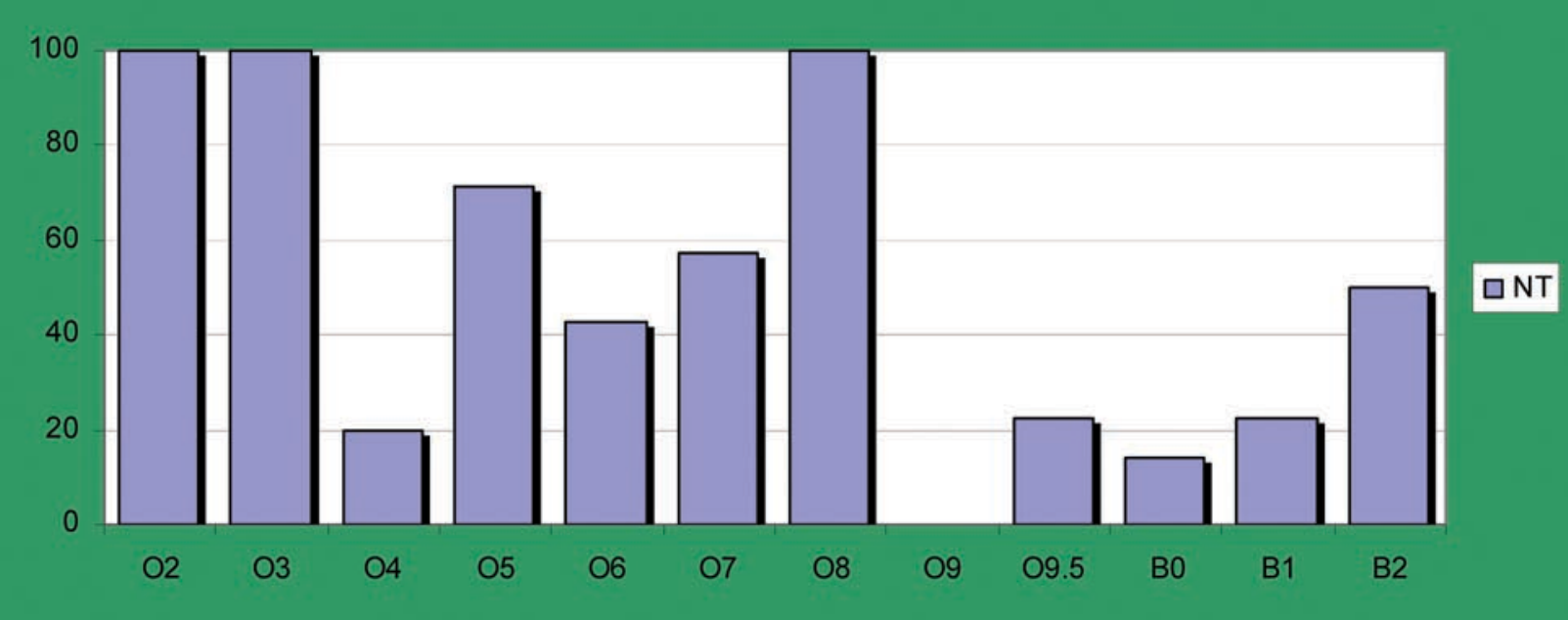}
\end{center}
\caption{Percentage of detected stars with spectral index flat or negative [NT and (NT) in Table 1], as a function of spectral type. The lower values obtained for spectral types O4 and O9 are discussed in the text.}\label{fig3}
\end{figure}

\begin{figure}[!ht]
\begin{center}
\includegraphics[bb= 0 0 438 212,width=11cm]{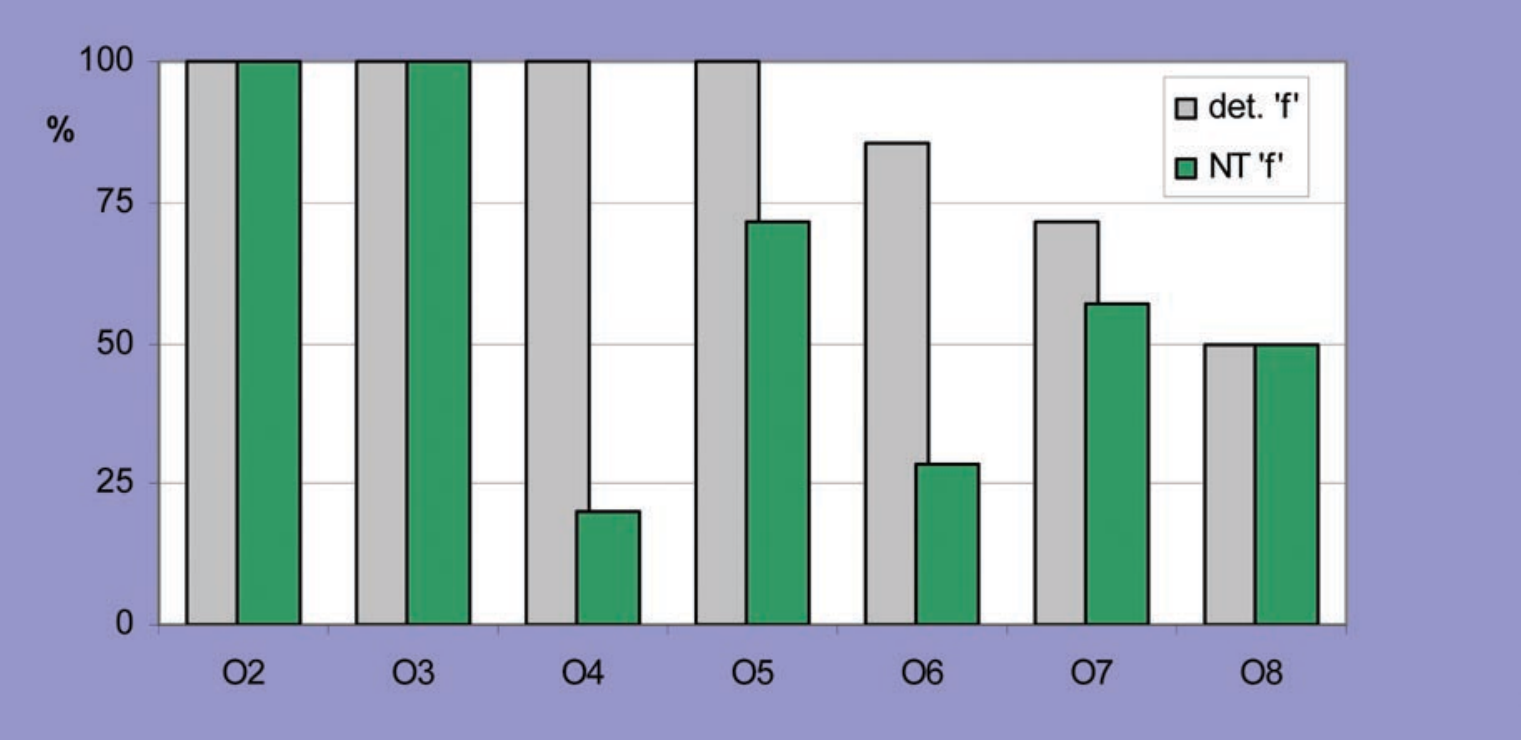}
\end{center}
\caption{Gray bars: percentage of ´f´-tagged (detected) stars; green bars: percentage of NT ´f´-tagged stars, both as a function of spectral type.}\label{fig4}
\end{figure}
\vspace{1.5cm}

\begin{figure}[!ht]
\begin{center}
\includegraphics[bb=0 0 425 201,width=11cm]{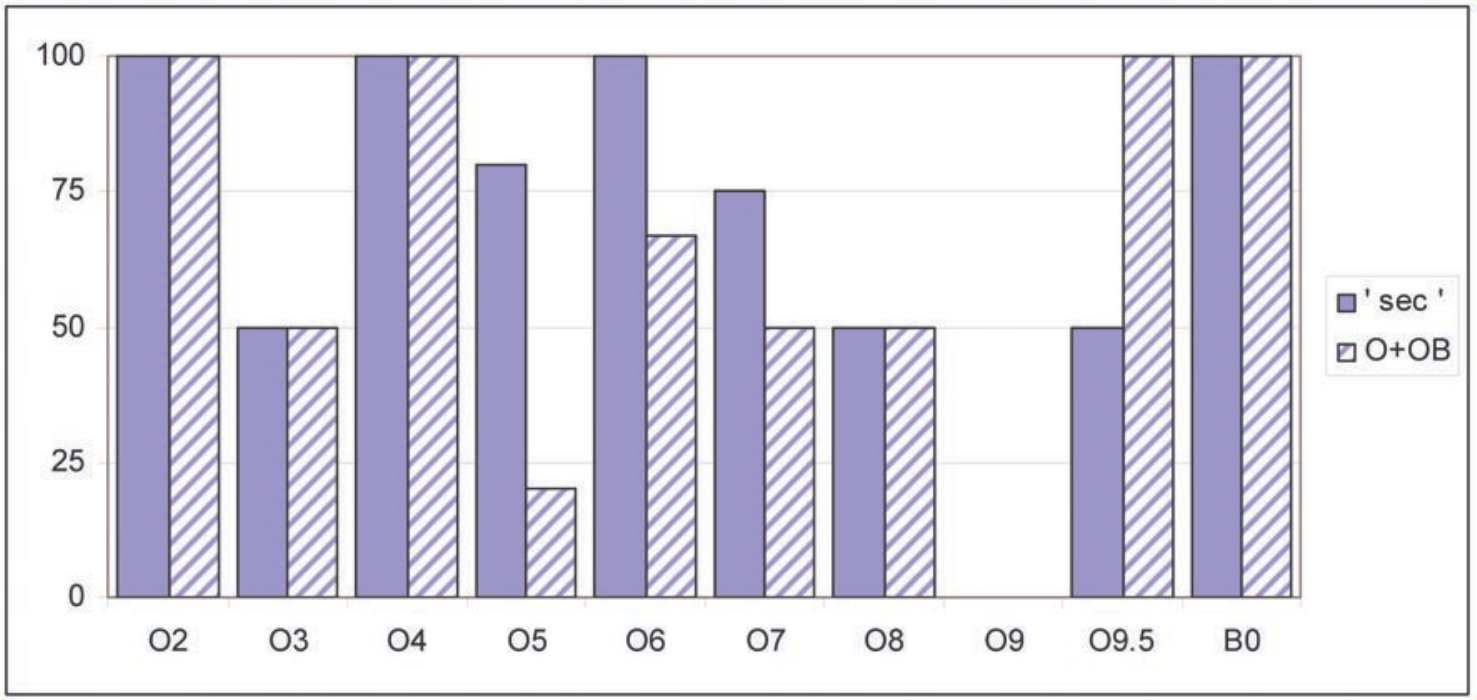}
\end{center}
\caption{Percentage of NT stars that have a (secondary) companion, of any spectral type (in purple), compared with the ones that belong to an O+OB system (hatched purple).}\label{fig5}
\end{figure}

\begin{figure}[!ht]
\begin{center}
\includegraphics[bb=0 0 181 159,width=8cm]{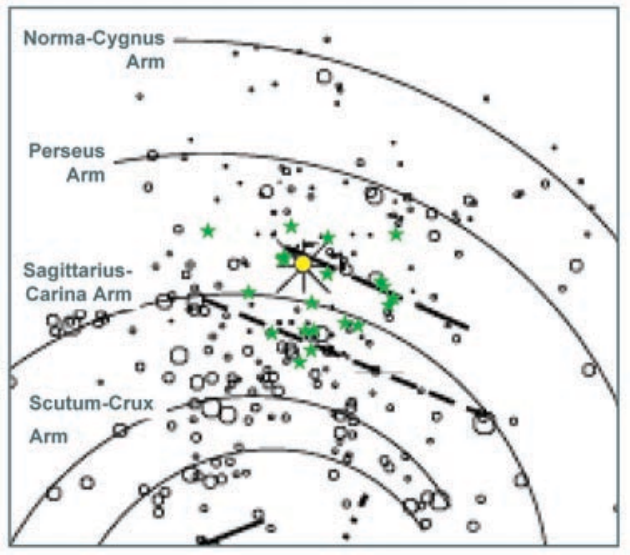}
\end{center}
\caption{Location of the detected non-thermal OB stars (green stars) at the galactic plane (spiral arm pattern and locations of HII regions -black open circles- from Russeil 2003). The yellow circle represents the Sun.}\label{fig6}
\end{figure}

Besides the issues enumerated in Sect. 4.1, there is a very strong constraint when performing a statistical analysis from the data of Table 1: the sample is still {\em very small}. Consequently, the results that can be drawn from the analysis must be taken as indicative in the low-significance limit. The results, rather than providing certainties, open many questions.\\

Figure 3 presents the percentage of NT stars vs spectral type. 
It is noteworthy that out of five O4 stars, the one classified as NT is the only known to be a binary. Why all O4 ``single'' (i.e. not yet known as in a binary or multiple system) are thermal emitters? Four O9 stars have been detected, though at only one frequency: is it a coincidence that no non-thermal O9 have been detected?\\

Figure 4 displays information on the `f' sub-classification, characteristic of evolved O prior to the WR stage. 
About 22 stars with spectral types O2 -- O6 were detected, all tagged with `f' (see Ma\'{\i}z Apell\'aniz et al. 2004 for a comprehensive classification of Of stars). Why none no-Of star up to O6 has been detected? Altogether, 25 stars are classified as f': eight as ((f)), two as (f) and 15 as f. Does this say something about the lifetime of each `f' sub-phase?\\

Information on the binarity status of the detected stars is presented in Fig. 5. Massive secondary companions have been identified in a couple of cases by visual observations (e.g. HD 93129A). Optical spectroscopic analysis and speckle interferometry permitted to find the spectral class of the secondary, or at least the type of system (eclipsing, spectroscopic, etc., see Mason et al. 2009).
Are all NT emitters in massive binary systems? How is it possible to establish the existence of ONE single NT emitter (e.g. HD 168112)? A detailed study on binarity (visual, spectroscopic, eclipsing types) should be performed at least on the detected sample.\\

The spectral index derived from the observations ranges within the interval ($-2.4$, 1.3): is there any consistent model to explain such a broad spectral index range? How accurate are the error estimates in flux densities? Are we able to state that if the mass loss rate derived from observations is larger than the expected one ($dM/dt_{\rm obs}$ $>$ $dM/dt_{\rm exp}$) then there is contribution of non thermal emission? \\

\section{Prospects in the light of forthcoming observational facilities}

The detected OB stars at radio waves are more than 60. Unfortunately, this is a very low number when studying galactic OB stars. Moreover, there is a flux limit on detection, strongly affected by the stellar distance. Figure 6 shows the distribution of the sources labeled as NT and (NT) in Table 1, on the local galactic environment: no pattern can be observed. These sources are closer than 3 kpc. Currently there is a limiting factor for detection at larger distances, and this is the telescope observing time and achievable minimum noise. Radio telescopes with better sensitivity and angular resolution are fundamental to build a more homogeneous sample.\\ 

The Expanded Very Large Array, and the e-Merlin\footnote{See www.aoc.nrao.edu/evla/astro/,  www.merlin.ac.uk/e-merlin/.} upgrade will be operational soon, in 2009. In the southern hemisphere, the Australia Telescope Compact Array has very recently been updated with new receivers: the CABB (or Compact Array Broadband Backend, www.narrabri.atnf.csiro.au/observing/). Total bandwidths of 128 MHz are replaced with 2 GHz ones, improving sensitivity in a factor 4. The new system is available for use from the June 2009 semester.\\

In the near future powerful instruments will be working along the whole radio spectrum: from a few MHz (cm) to THz (sub-mm) emission will be
probed. The Australia Square Kilometre Array Pathfinder ({\bf ASKAP}) is a 36 12m-antenna interferometer with a 30\,deg$^2$ field-of-view, at present being built in Western Australia. This instrument will observe, in principle, up to 2 GHz, with an angular resolution of some arcsecs, and is supposed to be fully operational in 2013. The Square Kilometre Array (SKA) is planned with about 180 antennas, observing frequencies from 0.1 to 25 GHz, beginning at the end of next decade. These telescopes will be able to discover large numbers of NT stellar systems.\\

The Atacama Large Millimeter Array ({\bf ALMA}) will be a mm+sub-mm array facility, built by a consortium from Europe, USA, Japan, and Canada, among other partners. It will include 50 dishes of 12 m plus a central, more compact, array (12 7m-antennas and 4 12m-antennas). The observing frequencies will cover from 30 to 720 GHz. The expected angular resolution will be that of the Next Generation Space Telescopes (JWST, SKA, ELT), and will reach tenth of arcsecs. The instrument is due to 2012. Contributions by ALMA to the study of early-type stars could be the detection of (thermal) fluxes of massive stars, and a meticulous study about mass loss rates; the mapping of the wind structure with unprecedented high dynamic range and angular resolution of 0.1''; the imaging of the gas kinematics and distribution in stellar environments, and the study of formation regions of massive protostars. All this conforms a promising future.

\vspace{0.5cm}
\acknowledgements 
P.B. greatly acknowledges the invitation from the SOC of the HEPIMS Workshop, and all the support including funding provided by the LOC. Many thanks especially to Dr. Josep Mart\'{\i} and Dr. Pedro L. Luque Escamilla. P.B. is indebted to Gustavo E. Romero for his help in all steps involved in the preparation of the present review, and wants to thank also Josep Mar\'{\i}a Paredes and Sean M. Dougherty. This work was partially supported by the Argentine agencies CONICET (PIP 452 5375) and ANPCyT (PICT-2007-00848).


\begin{thebibliography}{}
\bibitem[]{}Abbott, D. C., Bieging, J. H., Churchwell, E., \& Cassinelli, J.P. 1980, ApJ, 238, 196
\bibitem[]{}Abbott, D. C., Bieging, J. H., \& Churchwell, E. 1984, ApJ, 280, 671
\bibitem[]{}Benaglia, P., Cappa, C. E., \& Koribalski, B. S. 2001, A\&A, 372, 952
\bibitem[]{}Benaglia, P., \& Koribalski, B. 2004, A\&A, 416, 171
\bibitem[]{}Benaglia, P., Romero, G. E., Koribalski, B., \& Pollock, A. M. T. 2005, A\&A, 440, 743
\bibitem[]{}Benaglia, P., Koribalski, B., \& Albacete Colombo, J. F. 2006, PASA, 23, 50
\bibitem[]{}Benaglia, P., \& Koribalski, B. 2007, in ``Massive Stars in
Interacting Binaries'', A. F. J. Moffat, \& N. St-Louis (eds.), ASP Conf. Ser. 367, p. 179
\bibitem[]{}Benaglia, P., Vink, J. S., Mart\'{\i}, J., et al. 2007, A\&A, 467, 1265
\bibitem[]{}Bieging, J. H., Abbott, D. C., \& Churchwell, E. B. 1989, ApJ, 340, 518
\bibitem[]{}Biermann, L. 1951, Zs. f. Astrophys. 29, 274
\bibitem[]{}Blomme, R., Prinja, R. K., Runacres, M. C., \& Colley, S. 2002, A\&A, 382, 921
\bibitem[]{}Blomme, R., van de Steene, G. C., Prinja, R. K., et al. 2003, A\&A, 408, 715
\bibitem[]{}Blomme, R., van Loo, S., De Becker, M., et al. 2005, A\&A, 436, 1033
\bibitem[]{}Blomme, R., De Becker, M., Runacres, M. C., et al. 2007, A\&A, 464, 701
\bibitem[]{}Bulut I., \& Demircan O. 2007, MNRAS, 378, 179 
\bibitem[]{}Campbell, W. W. 1892, A\&A, 11, 799
\bibitem[]{}Castor, J. I., Abbott, D. C., \& Klein, R. I. 1975, ApJ, 195, 157
\bibitem[]{}Contreras, M. E., Rodriguez, L. F., Tapia, M., et al. 1997, ApJ, 488, L153
\bibitem[]{}De Becker, M., Rauw, G., Blomme, R., et al. 2004, A\&A, 420, 1061
\bibitem[]{}De Becker, M., Rauw, G., Blomme, R., et al. 2005, A\&A, 437, 1029
\bibitem[]{}De Becker, M. 2007, ARA\&A, 14, 171
\bibitem[]{}Drake, S. A. 1990, AJ, 100, 572 
\bibitem[]{}Drake, S. A., Abbott, D. C., Bastian, T. S., et al. 1987, ApJ, 322, 902
\bibitem[]{}Dougherty, S. M. 1993, PhD Thesis Calgary University
\bibitem[]{}Dougherty, S. M., \& Kennedy, M. 2009 (in press)
\bibitem[]{}Gibson, D. M., \& Hjellming, R. M. 1974, PASP, 88, 652
\bibitem[]{}Howarth, I. D., \& Brown, A. 1991, in Proceedings of the IAU Symp. 143, K. A. van der Hucht \& B. Hidayat (eds.), p. 315
\bibitem[]{}Johnson, M. C. 1925, MNRAS, 85, 813
\bibitem[]{}Lamers, H. J. G. L. M., \& Leitherer, C. 1993, ApJ, 412, 771
\bibitem[]{}Leitherer, C., \& Robert, C. 1991 ApJ, 377, 629
\bibitem[]{}Leitherer, C., Chapman, J. M., \& Koribalski, B. 1995, ApJ, 450, 289
\bibitem[]{}Lucy, L. B., \& Solomon, P. M. 1970, ApJ, 159, 879
\bibitem[]{}Mason, B. D., Hartkopf, W. I., Gies, D. R., et al. 2009, ApJ, 137, 3358
\bibitem[]{}Ma\'{\i}z-Apell\'aniz, J., Walborn, N. R., Galu\'e, H. A., \& Wei, L. H. 2004, ApJS, 151, 103
\bibitem[]{}Milne, E. A. 1924, MNRAS, 84, 354
\bibitem[]{}Moran, J. P., Davis, R. J., Spencer, R. E., et al. 1989, Nature, 340, 449
\bibitem[]{}Morton, D. C. 1967, ApJ, 147, 1017
\bibitem[]{}Panagia, N., \& Felli, M. 1975, A\&A, 39, 1
\bibitem[]{}Parker, E. N. 1958, ApJ, 128, 664  
\bibitem[]{}Parker, E. N. 1960, ApJ, 132, 821
\bibitem[]{}Persi, P., Rodr\'{\i}guez, L. F., Tapia, M., et al. 1988, ASSL, 142, 227
\bibitem[]{}Petr-Gotzens, M. M., \& Massi, M. 2008, in ``Multiple Stars Across the H-R Diagram'', Springer-Verlag Berlin Heidelberg, p. 281
\bibitem[]{}Puls, J., Markova, N., Scuderi, S., et al. 2006, A\&A, 454, 652
\bibitem[]{}Rauw, G., Blomme, R., Waldron, W. L., et al. 2002, A\&A, 394, 993
\bibitem[]{}Romero, G. E., Benaglia, P., \& Torres, D. F. 1999, A\&A, 348, 868
\bibitem[]{}Schnerr, R. S., Rygl, K. L. J., van der Horst, A. J., et al. 2007, A\&A, 470, 1105
\bibitem[]{}Scuderi, S., Panagia, N., Stanghellini, et al. 1998, A\&A, 332, 251
\bibitem[]{}Setia Gunawan, D. Y. A., Chapman, J. M., Stevens, I. R., et al. 2003a, Proceedings of the IAU Symp. 212, K. A. van der Hucht, A. Herrero \& C. Esteban (eds.), p. 230
\bibitem[]{}Setia Gunawan, D. Y. A., de Bruyn, A. G., van der Hucht, K. A., \& Williams, P. M 2003b, ApJS, 149, 123
\bibitem[]{}Van Loo, S., Blomme, R., Dougherty, S. M., \& Runacres, M. C. 2008, A\&A, 483, 585
\bibitem[]{}Wendker, H. J. 1995, A\&AS, 109, 177
\bibitem[]{}Wolf, C. J. E., \& Rayet, G. A. P. 1867, Compte Rendu, 65, 291
\bibitem[]{}Wright, A.E., \& Barlow, M.J. 1975, MNRAS, 170, 41
\end{thebibliography}
\end{document}